\documentclass[footinbib,reprint,12pt,onecolumn,floatfix,amsmath,prl,aps,preprintnumbers]{revtex4}

\usepackage{graphics}
\DeclareGraphicsExtensions{.png,.pdf}
\usepackage{epsfig}
\usepackage{todonotes}
\usepackage{bm}% bold math
\usepackage{color}

\usepackage{makeidx}
\usepackage{hyperref}
\begin{document}

%-----------------------------------------------------------------
\def\Tau{\mbox{\boldmath $\tau$}}

\title{\bf Theory of magnetic ordering in the heavy rare earths: ab-initio electronic origin of pair- and four- spin interactions}

\author{Eduardo Mendive-Tapia}

\affiliation{Department of Physics, University of Warwick, Coventry CV4 7AL, U.K.}
\date{\today}

\author{Julie B. Staunton}

\affiliation{Department of Physics, University of Warwick, Coventry CV4 7AL, U.K.}
\date{\today}
%%%%%%%%%%%%%%%%%%%%%%%%%%%%%%%%%%%%%%%%%%%%%%%%%%%%%%%%%%%%%%%%%%%%%%%%%%%%%%%%%%%
\begin{abstract}
\textbf{
We describe an ab-initio disordered local moment theory for long period magnetic phases and investigate the temperature and 
magnetic field dependence of the magnetic states in the heavy rare earth elements (HRE), namely paramagnetic, conical 
and helical anti-ferromagnetic(HAFM), fan and ferromagnetic (FM) states. We obtain a generic HRE magnetic phase 
diagram which is consequent on the response of the common HRE valence electronic structure to f-electron magnetic moment 
ordering. The theory directly links the first order HAFM-FM transition to the loss of 
Fermi surface nesting as well as providing a template for analysing the other phases and exposing where f-electron 
correlation effects are particularly intricate. Gadolinium, for a range of hexagonal, close-packed lattice constants, $c$ and $a$, is the 
prototype and applications to other HREs are made straightforwardly by scaling the 
pair and quartic local moment interactions with de Gennes factors and choosing appropriate lanthanide contracted $c$ and $a$ values.
}
\end{abstract}

%\pacs{71.18.+y, 71.20.Eh, 75.10.-b, 75.10.Lp, 75.30.Kz}

\maketitle

%----------------------------------------------------------------------------------------------------------------------------------
%----------------------------------------------------------------------------------------------------------------------------------
%\tableofcontents
%\pagenumbering{arabic}
%----------------------------------------------------------------------------------------------------------------------------------
%----------------------------------------------------------------------------------------------------------------------------------
Close scrutiny and {\it ab-initio} description of the magnetism of rare earth materials is motivated by its increasing importance for many 
applications as well as the fundamental interest of the strongly-correlated f-electrons underpinning it. A benchtest for this task and an outstanding challenge in its 
own right is to explain the diverse magnetism of the heavy rare earth (HRE) elements. 

The lanthanides from gadolinium to lutetium order into an 
apparently complex array of magnetic phases~\cite{Mackintosh1} despite the common chemistry of their valence electronic structure (5d$^1$6s$^2$ 
atomic configuration). Under ambient conditions they crystallise into hexagonal close packed structures and the number of localized f-electrons per 
atom increases from seven for Gd's half-filled shell through to Lu's complete set of fourteen which causes the lanthanide contraction of the 
lattice~\cite{GschneidnerB}. The magnetism is complicated. When cooled through $T_c$, Gd's paramagnetic (PM) phase 
undergoes a second order transition to a ferromagnetic (FM) state whereas, at $T_N$, Tb, Dy and Ho form incommensurate,
helical antiferromagnetic (HAFM) phases where the magnetization spirals around the crystal c-axis. When the temperature is lowered further both Tb and 
Dy undergo a first order transition at $T_t$ to a FM phase with basal plane orientation and Ho forms a conical HAFM ground state. 

Further exotic phases 
emerge when the metals are subjected to magnetic fields and they have been extensively studied in 
experiments~\cite{Tb1,Herz1,Dy6,Dy7,Chernyshov1,Ho2,Cowley1,Ho3}. Below its $T_c$ Gd preserves its FM order as the strength of the magnetic 
field applied along its easy axis is increased. This is in sharp contrast to Tb~\cite{Tb7,Tb10,Tb11,Tb1}, Dy~\cite{Herz1,Dy6,Dy7,Chernyshov1,Dy8} and Ho~\cite{Ho2,Ho3} 
above $T_t$, which first distort their HAFM order (dis-HAFM) before undergoing a first order transition into a fan magnetic structure followed 
by a second order transition to a FM state with further increase in the magnetic field. Dy and Ho also exhibit signs of additional spin-flip and vortex 
transitions associated with subtle changes in measured magnetization curves. In this letter we argue that much of this diversity
stems directly from the valence electronic structure that all the HRE elements share.

%%%%%%%%%%%%%%%%%%%%%%%%%%%%%%%%%%%%%%%%%%%%%%%%%%%%%%%%%%%%%%%%%%%%
%%%%%%%%%%%%%%%%%%%%%%%FIGUREPD%%%%%%%%%%%%%%%%%%%%%%%%%%%%%%%%%%%%%
\begin{figure}[ht]
%\centering
%\hspace*{-0.5cm}
%\includegraphics[clip,scale=0.35]{DiagramaDy.eps}
\centerline{
\includegraphics[clip,scale=0.7]{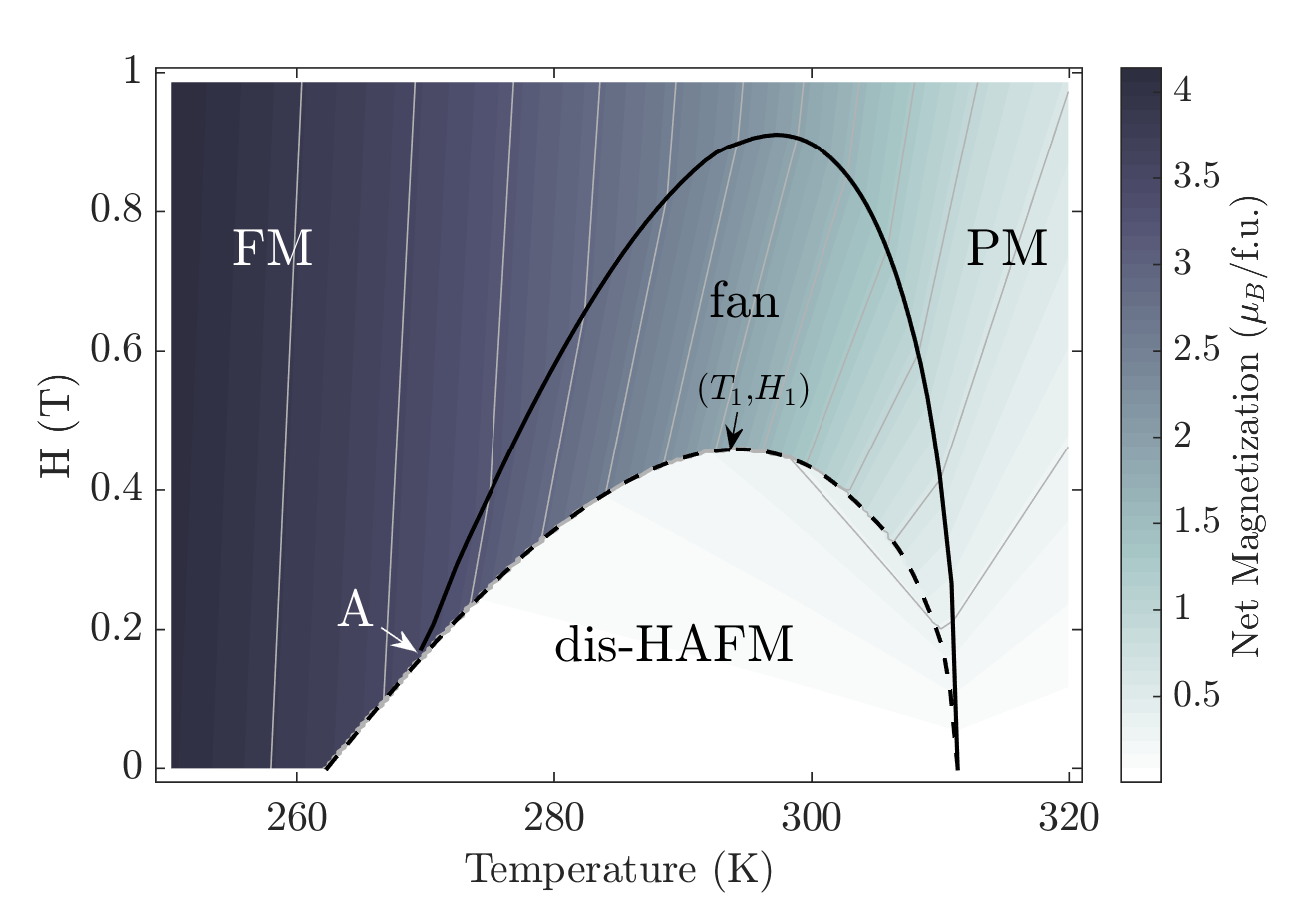}
}
\caption{(Color online) The generic magnetic $T$-$H$ phase diagram for a heavy lanthanide metal for $\textbf{H}$ applied
along the easy direction constructed from theory. Continuous (discontinuous) lines correspond to
second (first) order phase transitions and a tricritical point is marked (`A'). $c$ and $a$ lattice constants appropriate to Dy were used.}
\label{FIGPD}
\end{figure}
%%%%%%%%%%%%%%%%%%%%%%%%%%%%%%%%%%%%%%%%%%%%%%%%%%%%%%%%%%%%%%%%%%%%
 
Extensive experimental and theoretical investigations satisfactorily explain the onset of magnetic order from the PM state \cite{Evenson2,Andrianov4,Palmer1,Hughes1,Dobrich1,Andrianov1,Oroszlany1} via a detailed version of the famous Ruderman-Kittel-Kasuya-Yoshida (RKKY) 
pairwise interaction. The existence of nesting vectors $\textbf{q}_\text{nest}$ separating  parallel Fermi surface (FS) 
sheets of the valence electrons provokes a 
singularity in the conduction-electron susceptibility. This feature results in a $\textbf{q}_\text{nest}$-modulated 
magnetic phase~\cite{Mackintosh1,LiuBook}, identified as a HAFM structure, incommensurate with the underlying lattice. 
The lanthanide contraction changes the FS topology~\cite{Dugdale1} and acts as the decisive factor for the emergence of the nesting 
vectors. This has resulted in the construction of a universal crystallomagnetic phase diagram which links the magnetic ordering that emerges from the PM state 
to the specific $c$ and $a$ lattice parameters of a heavy lanthanide system~\cite{Hughes1,Andrianov4}.

The prominence of RKKY interactions in the discussion of lanthanide magnetism promotes a deeper inspection of the common HRE valence electronic structure. As magnetic order among the local f-electron moments of the HREs develops with decreasing temperature and/or strengthening applied magnetic field the valence electron glue spin-polarizes and qualitatively changes. An indication of this was found by Khmelevskyi et al. who calculated effective exchange interactions from the FM state to be different from those in the PM state~\cite{Khmelevskyi1}. This effect has a potentially profound feedback on the interactions between the magnetic moments and has wider relevance for other magnetic systems 
where the physics is also typically couched in RKKY terms, including giant 
magnetoresistive nanostructures~\cite{Zhou1}, rare earth clusters~\cite{Peters1}, magnetic semiconductors~\cite{HOhno1} and spin glasses~\cite{spinglass}.  In this letter we show how the response and feedback from the heavy 
lanthanide valence electrons to the ordering of local magnetic moments create multi-site interactions and determine the main features of the magnetic 
phase diagrams. We establish a reference, summarised in Fig.\ \ref{FIGPD}, against which these magnetic properties can be analysed to discriminate 
specific, subtle f-electron features. 

A simple classical spin model with pairwise exchange 
interactions, magnetic anisotropy contribution, and a Zeeman external magnetic field term describes magnetic field-driven phase transitions in some 
anti-ferromagnetic insulators~\cite{Nagamiya1,Nagamiya2}. If the local moments of a HAFM state are pinned by anisotropy and crystal field effects to spiral around a particular 
direction, the effect of a magnetic field causes a first order transition to a fan or conical phase where the spins now oscillate about the field 
direction. In higher fields, the fan or cone angles smoothly decrease to zero in the FM state. Such a model applied to the HREs addresses only part of 
phenomenon, however, since it fails to reproduce the first order dis-HAFM to FM and second order fan to FM transitions at low 
temperatures and fields. It misses a tricritical point in consequence.  In the seminal work by Jensen and Mackintosh \cite{Jensen1} where the 
formation of field-induced fan and helifan phases was investigated theoretically for the first time, the key aspects of the HRE magnetic phase diagrams 
were only reproduced if ad hoc temperature dependent pair-wise exchange interactions were incorporated from a fit to spin wave measurements conducted at 
a series of temperatures. We find instead that much of the magnetic phase complexity is directly traced back to the behavior of the valence
electrons. Evenson and Liu~\cite{Evenson2} maintained that the first order HAFM-FM transition is driven by a magnetoelastic
effect. We show rather that while there is a magnetostructural coupling it is not necessary for the transition.

We have extended the {\it ab-initio} density functional theory (DFT) based, disordered local moment (DLM) approach~\cite{Gyorffy1} to address this 
issue and construct a generic $H$-$T$ magnetic phase diagram of the HREs. Gd is a 
convenient prototype system owing to its seven localized f-electrons per atom in an S-state which form a large moment and the small crystal field and spin-orbit coupling effects that are prevalent. Choosing Gd enables us to abstract the common HRE valence electron effects 
on the magnetic properties and selecting the $c$ and $a$ lattice constants for other elements makes the analysis appropriate to Tb, Dy, Ho etc.  Fig.\ref{FIGPD} shows the results for Gd using the lattice parameters appropriate to 
Dy~\cite{Hughes1}.

The DLM-DFT describes the effects of thermally induced 'local moment' fluctuations on the underlying valence electronic structure of a magnet. For many materials such as the HRE's these magnetic excitations can be modeled by allowing the orientations of local, in the case of the HRE's f-electron, moments to vary very slowly on the time-scale of the valence electronic motions.  By taking appropriate ensemble averages over their orientational configurations DLM-DFT determines the system's magnetic properties and describes magnetic phase diagrams {\it ab-initio}~\cite{OnsagerCavity,Staunton5,Staunton7}, 
temperature dependent magnetic anisotropy~\cite{Staunton2,Staunton4,Staunton8} and field and temperature induced metamagnetic transitions~\cite{Staunton9}.
%Strong f-electron correlations are included via the self-interaction correction (SIC)~\cite{LudersSIC,Perdew1,Hughes1}. DLM-DFT's predictive modelling capability has been demonstrated by the recent theory/experimental study of Gd-Mg, Gd-Zn and Gd-Cd compounds~\cite{Staunton7}. 

The theory has the advantage that valence electronic structure can be monitored as a function of local moment disorder. This is highly pertinent owing to the recent development of advanced time-dependent spectroscopy techniques. Time-resolved resonant X-ray and 
ultrafast magneto-optical Kerr studies confirm the central tenet that the dynamics of the HRE core-like f-electrons are on a much longer time-scale than the excitations of the valence electrons~\cite{Langner1}. Time- and angle-resolved photoemission (ARPES) studies have 
demonstrated the differing dynamics of spin-polarized valence states in correlated materials~\cite{Teichmann1,Andres1,Krieger1}.
%%%%%%%%%%%%%%%%%%%%%%%%%%%%%%%%%%%%%%%%%%%%%%%%%%%%%%%%%%%%%%%%%%%%
%%%%%%%%%%%%%%%%%%%%%%%FIGUREFS%%%%%%%%%%%%%%%%%%%%%%%%%%%%%%%%%%%%%
\begin{figure}
\centering
\includegraphics[width=150mm,height=50mm]{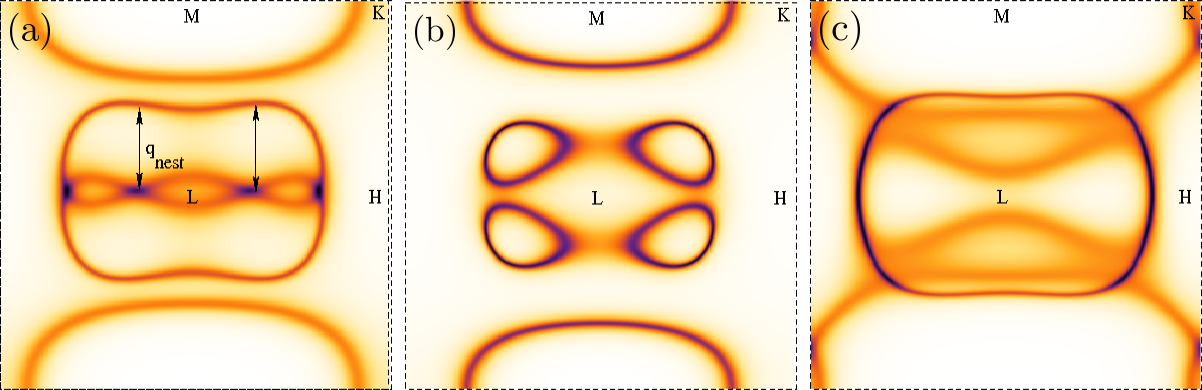}
%\includegraphics[clip,scale=0.17]{FigureFS.eps}
%\centerline{
%\includegraphics[width=28mm,height=30mm]{FigRE-2a.png}
%\includegraphics[width=28mm,height=30mm]{FigRE-2b-up.png}
%\includegraphics[width=28mm,height=30mm]{FigRE-2b-dn.png}
%\includegraphics[width=45mm,height=30mm]{FigRE-2a.png}
%\includegraphics[width=45mm,height=30mm]{FigRE-2b-up.png}
%\includegraphics[width=45mm,height=30mm]{FigRE-2b-dn.png}
%}
\caption{(Color online) The Bloch spectral function in the $LMHK$ plane at the Fermi energy for Gd with Dy's lattice constants for (a) the PM 
state and resolved into (b) majority spin and (c) minority spin components when there is an overall net average magnetization of 54\%, ($m_{FM}=0.54$), of 
the $T=$0K saturation value in the FM state. This is the value in our calculations (Fig.\ref{FIGPD}) in the FM phase just below the temperature $T_t$ of the HAFM-FM first order transition. 
${\bf q}_\text{nest}$ indicates the nesting wave-vector of the FS of the PM state and the shading represents the broadening from thermally induced local moment disorder.}
\label{FIGFS}
\end{figure}
%\todo[size=\tiny]{Replace Figure by updated one}
%%%%%%%%%%%%%%%%%%%%%%%%%%%%%%%%%%%%%%%%%%%%%%%%%%%%%%%%%%%%%%%%%%%%
%%%%%%%%%%%%%%%%%%%%%%%%%%%%%%%%%%%%%%%%%%%%%%%%%%%%%%%%%%%%%%%%%%%%

Fig.\ref{FIGFS} shows our calculated FS of Gd with Dy's lattice attributes within the DLM picture. Fig.\ref{FIGFS}(a) shows the FS when the moments are randomly oriented in the paramagnetic state. The 
nesting vectors responsible for the onset of Dy's HAFM state below $T_N$ are clearly seen~\cite{Hughes1}. Figs.\ref{FIGFS}(b) and (c) show the FS where now the 
local f-electron moments are oriented on average to produce an overall net average magnetization of 54\% of the $T=$0K saturation value. 
The FS is spin-polarized and neither majority nor minority spin component continues to show nesting. This dramatic change of FS topology hints at the valence electron's role in the HAFM-FM metamagnetic transition. It also concurs with conclusions drawn from D\"{o}brich \textit{et al.}'s~\cite{Dobrich1} angle-resolved photoemission measurements pointing to the magnetic exchange splitting of the FS as the 
principal mechanism for the fading of the nesting vectors~\cite{LiuBook} resulting in the stability of the FM phase at the ground state in Tb and Dy. 

To follow the repercussions of this insight we specify a generalised Grand Potential $\Omega ( \{\hat{e}_{n,i}\})$ from DFT in which the local moments 
are constrained to point along directions $\{\hat{e}_{n,i}\}$~\cite{Gyorffy1}. This quantity is averaged over many such configurations with a probability 
distribution $P \{\hat{e}_{n,i}\}= \prod_n \prod_i P_n (\hat{e}_{n,i})$ where $
P_n (\hat{e}_{n,i})= \frac{\exp ({\bf A}_n \cdot \hat{e}_{n,i})}{\int \exp ({\bf A}_n \cdot \hat{e}^{'}_{n,i})\,
d \hat{e}^{'}_{n,i}}$.
$n$ and $i$ count over layers stacked along the c-axis (i.e. the z-axis) and sites within a layer respectively. The local average, $\textbf{m}_n = 
\left<\hat{e}_{n,i}\right> = \large(-\frac{1}{A_n} +  \coth A_n \large) \hat{A}$, therefore defines an order 
parameter prescribed by the input $\{ {\bf A}_n \}$ values. A magnetically ordered  state
is specified by the set $\{\textbf{m}_n\}$. The PM 
state corresponds to $\{\textbf{m}_n\}=\{\textbf{0}\}$ 
and FM state to $\{\textbf{m}_n\}=\{\textbf{m}_{\text{FM}}\}$. 
An HAFM phase modulated by the wave
vector $\textbf{q}_0=(0, 0, q_0)$ applies when 
$\textbf{m}_n=m_{\text{HAFM}}\left(\cos(\textbf{q}_0\cdot\textbf{R}_n),\sin(\textbf{q}_0\cdot\textbf{R}_n), 0\right)$,
where $\textbf{R}_n$ indicates the position of the $n$-th layer. The average $ \left<\Omega (\{\hat{e}_{n,i}\})\right>$ is consequently a 
function of the $\{\textbf{m}_n\}$ magnetic order parameters, $\bar{\Omega}(\{\textbf{m}_n\})$.  By repeating the calculation for many sets of 
$\{\textbf{m}_n\}$ (i.e. $\{ \textbf{A}_n\}$ choices) and careful analysis~\footnote{See Supplementary Material for further information about DLM theory, the construction of the magnetic phase diagram, detail about the DLM-DFT data and evaluation of the effective pair interactions, and the magnetic phase diagrams of Gd with the attributes of Tb, Dy, and Ho.}
we find the internal magnetic energy $\bar{\Omega}$ to fit very well the expression
\begin{equation}
%\begin{split}
%\bar{\Omega}=-\sum_{n,n'} & {\mathcal{J}_{nn'}}\,\textbf{m}_n\cdot\textbf{m}_{n'} \\
%             -\sum_{n,n',n'',n'''} & {\mathcal{K}_{nn',n''n'''}(\textbf{m}_n \cdot \textbf{m}_{n'})
%(\textbf{m}_{n''}\cdot\textbf{m}_{n'''})},
%\end{split}
\bar{\Omega}=-\sum_{n,n'}({\mathcal{J}_{nn'}}
             +\sum_{n'',n'''}{\mathcal{K}_{nn',n''n'''}\textbf{m}_{n''}\cdot\textbf{m}_{n'''}})\textbf{m}_n\cdot\textbf{m}_{n'}.
\label{EQOmega}
\end{equation}

$\mathcal{J}_{nn'}$'s are interpreted as pair-wise local moment interactions and the quartic coefficients $\mathcal{K}_{nn',n''n'''}$ arise from the 
mutual feedback
between local moment magnetic order and the spin polarized valence electrons illustrated in Fig.\ref{FIGFS}.
%The summation in Eq.\ref{EQOmega} goes through all the lattice layer space.
% and $\bar{\Omega}_0$ is a constant.

For a phase diagram such as Fig.\ref{FIGPD} we construct the Gibbs free energy $\mathcal{G}$ of the system from
\begin{equation}
%\mathcal{G}=\bar{\Omega}(\{\textbf{m}_n\})-\sum_{n}{\left[\mu \textbf{m}_{n}\cdot\textbf{H}+TS_n(\textbf{m}_{n})+F_u(\textbf{m}_n) \right]},
\mathcal{G}=\bar{\Omega}-\sum_{n}{\left[\mu \textbf{m}_{n}\cdot\textbf{H}+TS_n(\textbf{m}_{n})+F_u(\textbf{m}_n) \right]},
\label{EQF}
\end{equation}
where $T$ and $S_n(\textbf{m}_{n})= -k_\text{B}\,\int P_n (\hat{e}) \ln P_n (\hat{e}) \text{d} \hat{e} $ are the temperature and the magnetic entropy of the 
\textit{n}-th layer, respectively ($k_\text{B}$ being 
Boltzmann's constant). The second term couples the external magnetic field $\textbf{H}$ to the local 
magnetic moments each with magnitude $\mu$. The last term, $F_u(\textbf{m}_n)= F_0\left<(\hat{e}_{n,i} \cdot \hat{z})^2\right>$~\cite{CallenCallen1}, describes a uniaxial anisotropy 
with strength $F_0$~\cite{McEwen1} and fixes the easy axis. For selected $T$ and $\textbf{H}$ values, the Gibbs free energy $\mathcal{G}$ is evaluated for the values $\{\textbf{m}_n\}$ which minimize it, i.e. 
$\nabla_{\textbf{m}_i}\mathcal{G}(\textbf{m}_i,\textbf{h}_i;T)=\textbf{0}$. This is accomplished when
\begin{equation}
{\bf A}_n = -\beta \left( \nabla_{\textbf{m}_n}\bar{\Omega} - \nabla_{\textbf{m}_n} F_u(\textbf{m}_n) - \textbf{H} \right)=  \beta \textbf{h}_{n},
\label{Weiss}
\end{equation}
where $\beta=1/k_{\text{B}}T$. The $\textbf{h}_{n}$'s are therefore the Weiss fields for this mean field theory~\cite{Gyorffy1,Staunton5,Staunton6,Staunton7}.

We carried out DLM-DFT calculations~\cite{Note1} for Gd within this framework for the $c$ and $a$ lattice parameters appropriate to Gd, Tb, Dy and Ho and in each case calculated charge
and magnetization densities self-consistently for the PM state ($\{\textbf{m}_n =0 \}$). The self interaction correction was used to capture the strong correlations of the 
f-electrons~\cite{LudersSIC,Perdew1,Hughes1}. A local moment of $\mu \approx$7.3 $\mu_B$ established on each Gd site.
We then divided the hexagonal lattice into 10 layer stacks and specified identical sets of $\{\textbf{m}_n\}$ ($n=1,\cdots,10$) 
values for each stack to define the magnetic order parameter for each layer.
For each $c$ and $a$ pair, using the effective one electron PM potentials, we calculated 
the $\textbf{h}'_{n} = -\nabla_{\textbf{m}_n} \bar{\Omega}$ values for each $\{\textbf{m}_n\}$ set.  By thoroughly sampling the extensive the $\textbf{m}_n$ space we tested and established a method~\cite{Note1} to extract the $\mathcal{J}_{nn'}$ and $\mathcal{K}_{nn',n''n'''}$ 
constants of Eq.\ref{EQOmega}. We also checked that higher order terms were vanishingly small~\cite{Note1}. Fig.\ref{FIGPD} summarises the results. At a value $m_{FM}=$0.503 the system undergoes a first order transition from a HAFM to FM state in zero field at $T_t=$262K which correlates with the FS topological changes depicted in Fig.\ref{FIGFS}. When the 4 site $\mathcal{K}_{nn',n''n'''}$'s are neglected in Eq.\ref{EQOmega} the calculated phase diagram is very different~\cite{Note1} and a FM phase does not appear at low temperatures and fields. 
 
\begin{figure}[ht]
%\centering
\hspace*{-0.5cm}
\includegraphics[clip,scale=0.7]{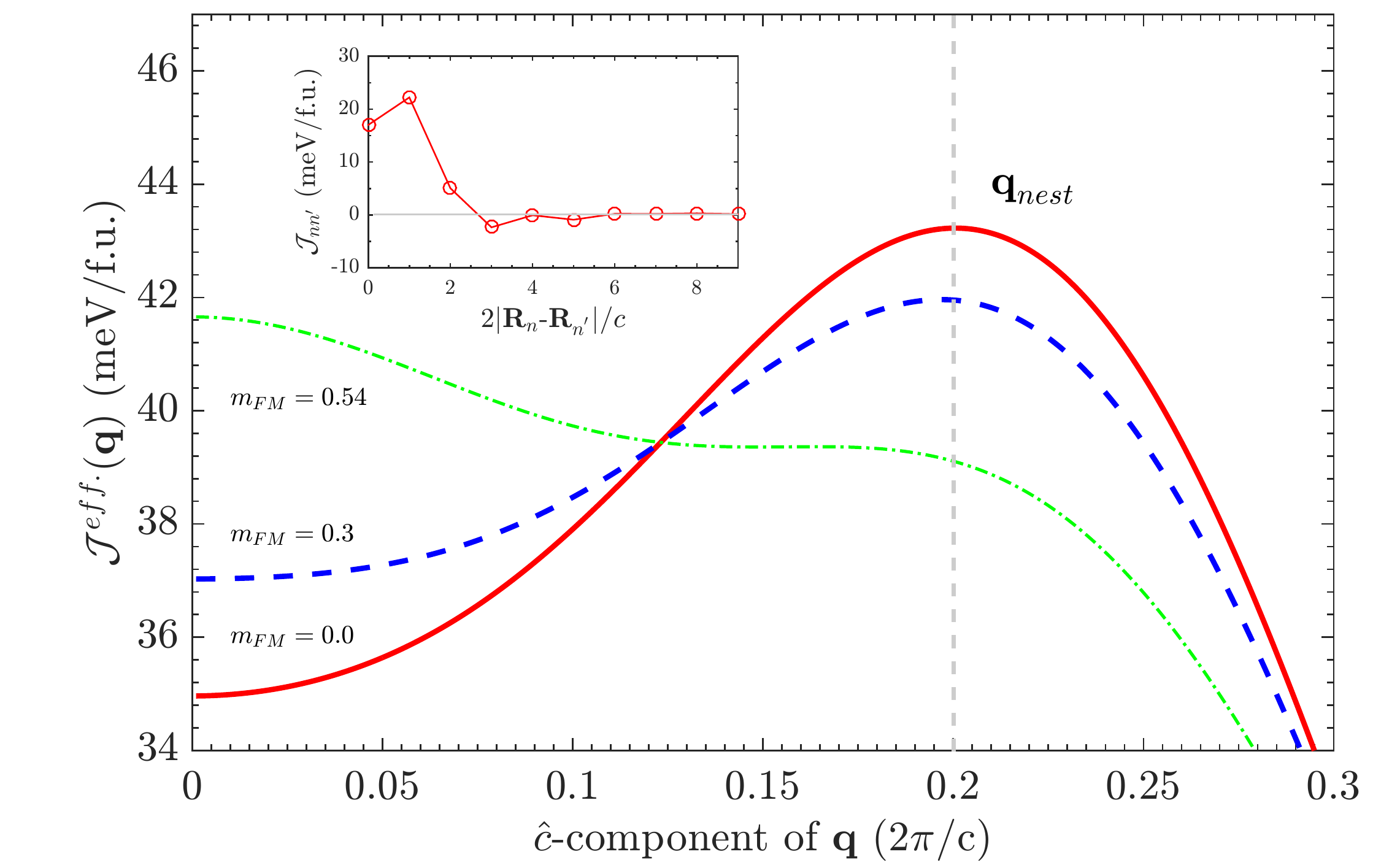}
\caption{(Color online) The lattice Fourier transform of the effective pair interactions 
$\mathcal{J}^{eff.}(\textbf{q})$ (red line) when $m_\text{FM}=0$ and its change when the FS is spin polarized, 
for finite $m_\text{FM}$ (dashed blue line for $m_\text{FM}=0.3$ and dot-dashed green line for $m_\text{FM}=0.54$).
The inset shows the dependence of the pair interactions $\mathcal{J}_{nn^{'}}$ on separation 
$R_{nn^{'}}=|{\bf R}_n -{\bf R}_{n^{'}}|$ for $m_\text{FM}$=0.}
\label{FIGopt}
\end{figure}

In the presence of long-ranged magnetic order, quantified by $m_\text{FM}$, effective
pair interactions mediated by the valence electrons affected by the long-range magnetic order can be specified as $\mathcal{J}^{eff.}_{nn^{'}} = \mathcal{J}_{nn^{'}} + \sum_{ n''n'''} \mathcal{K}_{nn',n''n'''} m^2_\text{FM}$~\cite{Note1} and they incorporate the influence of the 4-site terms from Eq.\ref{EQOmega}. Fig.\ref{FIGopt} shows their lattice Fourier transform relevant to Figs.\ref{FIGPD} and \ref{FIGFS} revealing the effect of the valence electron spin polarization. 
As shown in the inset, for $m_\text{FM}=0$, the interactions have a long-ranged oscillatory nature so that $\mathcal{J}^{eff.} ({\bf q})$ peaks at $\textbf{q}_\text{nest} \approx 
0.2 \frac{2\pi}{c}\hat{c}$, (full red line). This is a direct consequence of the FS nesting shown in Fig.\ref{FIGFS}(a) and which drives the HAFM magnetic order. We also show $\mathcal{J}^{eff.} ({\bf q})$ for non-zero $m_\text{FM}$. When $m_\text{FM}$=0.54 (green, dot-dashed lines), the value in the FM phase just below $T_t$, $\mathcal{J}^{eff.} ({\bf q})$ peaks at $\textbf{q}=\textbf{0}$ showing how the development of long-range magnetic order 
has favored the shift towards ferromagnetism. This confirms the role of the spin-polarized valence electrons and altered FS topology exemplified in Figs.\ref{FIGFS}(b) and (c). 

%%%%%%%%%%%%%%%%%%%%%%%%%%%%%%%%%%%%%%%%%%%%%%%%%%%%%%%%%%%%%%%%%%%%
%%%%%%%%%%%%%%%%%%%%%%%FIGUREopt%%%%%%%%%%%%%%%%%%%%%%%%%%%%%%%%%%%%

\begin{table*}[t]
\begin{center}
\hspace*{-0.6cm}
\begin{tabular}{ c || c c | c c | c c c c | c c | c }
 \hline\hline
 Element &
 $T_{\text{N}} (\text{K})$ & $T^{\text{exp}}_{\text{N}} (\text{K})$ &
 $T_{\text{t}} (\text{K})$ & $T^{\text{exp}}_{\text{t}} (\text{K})$ &
 $T_{\text{1}} (\text{K})$ & $T^{\text{exp}}_{\text{1}} (\text{K})$ &
 $H_{\text{1}} (\text{T})$ & $H^{\text{exp}}_{\text{1}} (\text{T})$ &
 $T_{\text{A}} (\text{K})$ & $H_{\text{A}} (\text{T})$ &
 References \\
 \hline
% Gd & 274 ($T_{\text{C}}$) & 293($T_{\text{C}}$) & -    & -            &   -   &  -   & -   &  -  &  -  &  -  & \cite{Mackintosh1} \\
 Tb & 214  & 229 & 206  & 222 & 211 & 224 to 226 & 0.03    & 0.02 to 0.03 & 207 & 0.01 & \cite{Tb7,Tb10,Tb11,Tb1} \\
 Dy & 145  & 180 & 90 & 90  & 129 & 165 to 172 & 0.43    & 1.1 to 1.2     & 94  & 0.07 & \cite{Herz1,Chernyshov1,Dy8} \\
 Ho & 94 & 133 & -    & 20  & 65  & 110        & 1.03    & 3.0            & -   & -     & \cite{Ho2,Ho1,Ho4} \\
 \hline\hline
\end{tabular}
\caption{Application of the theory to Tb, Dy and Ho and comparison with experiment. The values of $T_N$, $T_t$ and the $T$ for the highest \textbf{H} for the dis-HAFM phase are compared to experiment ($T_1$,$H_1$ in Fig.~\ref{FIGPD}). Theoretical estimate of the tricritical point (`A') is also given. Gd has a PM-to-FM second order transition at $T_C$=274K ($T_C$=293K in experiment~\cite{Mackintosh1}).}
\end{center}
\end{table*}

%To construct the magnetic phase diagram the minimization of Eq.\ref{EQF} was carried out and the free energy of a magnetic structure as function of $T$ and $\textbf{H}$ calculated.
By comparing Gibbs free energies of the FM, HAFM, conical, fan, and helifan structures obtained, we constructed the $T$-$H$ phase 
diagram. Fig.\ref{FIGPD} shows the results using $c$ and $a$ values appropriate to Dy when 
$\textbf{H}$ was applied along the easy direction and continuous/dashed lines correspond to second/first
order phase transitions. We imposed a single site uniaxial anisotropy of typical magnitude $F_0$=+6.3meV/site~\cite{McEwen1} which precluded the conical phase when the magnetic field was applied in the easy ab-plane~\cite{Note1}. 

The figure reproduces all the main features that the experimentally measured magnetic phase diagrams of heavy lanthanide metals and their alloys have in common. 
There is the first order HAFM-FM transition in the absence of \textbf{H} at $T_t$.
Then for increasing values of \textbf{H} applied along the easy direction the helical structure initially distorts before transforming to the fan structure. 
Increasing \textbf{H} further stabilizes the FM phase. There is, also in line with experimental findings, a second order transition from the fan to FM phase in finite field on cooling and we find a tricritical point which is marked `A' in Fig.\ \ref{FIGPD}.
%Fig.\ \ref{FIGPD}(b) shows that the HAFM structure transforms to a conical phase when a magnetic field is applied along the hard axis. The FM-cone phase transition is of first(second)-order at lower(higher) temperatures, as indicated in the figure and consistent with experiment \cite{Chernyshov1}.
%The phase diagram is radically altered and qualitatively at odds with experimental results~\cite{Note1} if we neglect the quartic terms, i.e. set $\mathcal{K}_{nn^{'},n^{''}n^{'''}}=0$, and so omit the feedback between the valence electronic structure and lanthanide local f-electron magnetic moment order.  

We can adapt this Gd-prototype model to a specific heavy lanthanide element or alloy by using suitable lanthanide-contracted lattice constant values~\cite{Hughes1} and
accounting for the specific f-electron configuration. The Gd ion has orbital angular momentum $L=0$ and
negligible spin orbit coupling effects. LS-coupling, however, is important for the HREs. A simple measure to account for the different total angular momentum values, $J$, is to scale the $\mathcal{J}_{nn'}$ and $\mathcal{K}_{nn',n''n'''}$ interactions with the famous de Gennes factor~\cite{Mackintosh1} $(g_J - 1)2J(J + 1)$, where $g_J$ is 
the Lande g-factor. We applied this treatment to Gd, Tb, Dy and Ho and each metal except Gd had a phase diagram of the form of Fig.~\ref{FIGPD} 
consistent with experiment.

In Table I we compare results with those available from experiment for $T_N$ and $T_t$, and the values of \textbf{H} for the highest $T$ for the dis-HAFM phase. We also give a theoretical estimate of the tricritical point. The comparison overall shows that the theory correctly captures trends and transition temperature and field magnitudes. When the $c$
and $a$ values are further decreased our model predicts that the 
FM phase does not appear at low $T$ and fields in accord with some experiments~\cite{Edwards1}. Discrepancies between the model and experiment can 
further highlight where f-electron correlation~\cite{Locht1} effects are leading to more complicated physics. For example the complex spin slip phases in Ho reported at low temperatures~\cite{Ho2,Ho3} are not found in our model.  The same applies to the vortex and helifan phases inferred from some experimental studies~\cite{Chernyshov1,Jensen1}.

In summary we claim a dominant role for the valence electrons in the temperature-field magnetic phase diagrams of the HRE metals. Our ab-initio theory incorporates lattice structural effects coming from the lanthanide contraction on this glue and makes the link between the changing topology of the FS, observed experimentally, and the evolving long range magnetic order of the f-electron moments which triggers the first order transition between HAFM and FM states. Tricritical points are also predicted. This generic valence electron effect produces pair-wise and four-site interactions among the localized f-electron moments and rules out the necessity to invoke ad-hoc temperature 
dependent effective interactions or magnetostrictive effects. A simple de Gennes factor scaling of the 
interactions along with a phenomenological measure of magnetocrystalline anisotropy to fix the easy magnetization plane enables this model to be applied broadly to HREs. We propose the model as a filter to identify subtle lanthanide f-electron correlation effects for further scrutiny.

\begin{acknowledgments}
The authors gratefully acknowledge discussions with O. Trushkevych and 
R. S. Edwards. The work was supported by EPSRC (UK) grants  EP/J06750/1 and EP/M028941/1.
\end{acknowledgments}

\section{Theory of magnetic ordering in the heavy rare earths: ab-initio electronic origin of pair- and four- spin interactions - Supplementary Information}

\subsection{Theory for the \textit{Ab-initio} Gibbs Free Energy}

In the manuscript we implemented Density Functional Theory (DFT) -based Disordered Local Moment (DLM) theory~\cite{Gyorffy1,Staunton5,Staunton7,Hughes1} to evaluate the interplay between localized magnetic moments, associated with the f-electrons, and the underlying valence electronic structure in the heavy rare earth (HRE) elements. 
%If the orientations of these disordered local moments (DLMs) vary slowly when compared to the fast itinerant electronic motions, their thermal fluctuations can be incorporated to examine the effect of magnetic order on the valence electronic structure. Under these circumstances the partition function of the system can be constrained to the orientational configuration $\{\hat{e}_{n,i}\}$ of the DLMs. A probability distribution $P\{\hat{e}_{n,i}\}$ of a given configuration is naturally defined in this approach. 
As explained in the manuscript, we specify a `generalised' constrained Grand potential $\Omega(\{\hat{e}_{n,i}\})$ that is then averaged over all the DLM configurations giving rise to the internal energy $\bar{\Omega}(\{\textbf{m}_n\})=\left<\Omega(\{\hat{e}_{n,i}\})\right>$. The single site probabilities $P_n(\hat{e}_{n,i})=\frac{\exp ({\bf A}_n \cdot \hat{e}_{n,i})}{\int \exp ({\bf A}_n \cdot \hat{e}^{'}_{n,i})\,d \hat{e}^{'}_{n,i}}$ are estimated within a mean-field approach in terms of the Weiss fields $\textbf{A}_{n}$, given in Eq.\ (3) in the manuscript.
%We treat with a `spin-only' mean-field Hamiltonian governing the orientations of the local magnetic moments. \textit{Ab initio} machinery is used, therefore, to capture the influence of the magnetic ordering at finite temperatures. 

The natural quantities from this approach are the order parameters where
\begin{equation}
\textbf{m}_n = \left<\hat{e}_{n,i}\right> = \large(-\frac{1}{A_n} + {\it \coth} A_n \large) \hat{A}
\label{OP}
\end{equation}
%$\{\textbf{m}_n = \left<\hat{e}_{n,i}\right> = \large(-\frac{1}{A_n} + {\it \coth} A_n \large) \hat{A}\}$
describes the magnetic order within the n-th ferromagnetic layer. We express the free energy of the system as $\mathcal{F}=\bar{\Omega}-Tk_\text{B}\sum_n S_n(\textbf{m}_{n})$, where $k_\text{B}$ is the Boltzmann constant and the entropy per site in the n-th ferromagnetic layer is calculated as $S_n=1+\ln(4\pi)+\ln\left(\sinh(A_n)/A_n\right)-A_n\coth(A_n)$. The Gibbs free energy $\mathcal{G}$ is then obtained by adding the external magnetic field Legendre transformation to the free energy $\mathcal{F}$ and a term to describe a uniaxial anisotropy, which leads to Eq.\ (2) in the manuscript. An analytical expression for the uniaxial anisotropy can be derived by performing the integral $F_u(\textbf{m}_n)= F_0 \left<(\hat{e}_{n,i} \cdot \hat{z})^2\right>=F_0 \int P_n(\hat{e}_{n,i})(\hat{e}_{n,i}\cdot\hat{z})^2\text{d}\hat{e}_{n,i}$ which becomes
\begin{equation}
%F_u(\textbf{m}_n)= F_0 \left[\cos^2\theta-\frac{1}{A_n}\left(\frac{-1}{A_n}+\coth A_n\right)(3\cos^2\theta-1)\right]
%F_u(\textbf{m}_n)= F_0 \left[\frac{(\textbf{A}_n\cdot\hat{z})^2}{A^2_n}-\frac{1}{A_n}\left(\frac{-1}{A_n}+\coth A_n\right)\left(3\frac{(\textbf{A}_n\cdot\hat{z})^2}{A^2_n}-1\right)\right]
F_u(\textbf{m}_n)= F_0 \left[(\hat{\textbf{A}}_n\cdot\hat{z})^2-\frac{1}{A_n}\left(\frac{-1}{A_n}+\coth A_n\right)\left(3(\hat{\textbf{A}}_n\cdot\hat{z})^2-1\right)\right],
\label{Fu}
\end{equation}
%%%%%%%%%%%%%%%%%%%%%%%%%%%%%%%%%%%%%%%%%%%%%%%%%%%%%%%%%%%%%%%%%%%%%%%%%%%%%%%%%%%
where $F_0$ is the strength of the anisotropy term. Prior to minimization of $\mathcal{G}$ we need to evaluate the dependence of the internal energy $\bar{\Omega}(\{\textbf{m}_n\})$ on the order parameters $\{\textbf{m}_n\}$ (or $\{\textbf{A}_n\}$). Our DLM-DFT theory can accomplish this as it explicitly calculates $\nabla_{\textbf{m}_n}\bar{\Omega}$~\cite{Gyorffy1}. This requires an extensive exploration of the directional derivative in order to obtain an accurate description of $\bar{\Omega}$. In the following section we explain in detail this methodology.

\subsubsection{Minimization of the Gibbs Free Energy}
\label{Method}

Our methodology to minimize $\mathcal{G}$ consists in solving Eq.\ (\ref{OP}) together with Eq.\ (3) in the manuscript. The equations are solved in an iterative process for given values of $\beta$ (or the temperature) and the magnetic field $\textbf{H}$. Firstly, an initial arrangement of $\{\textbf{A}^0_n\}$ is chosen. The corresponding values of $\{\textbf{m}^0_n\}$ are calculated from Eq.\ (\ref{OP}). Eq.\ (3) of the manuscript is used to evaluate $\{\textbf{A}_n\}$ again from these $\{\textbf{m}^0_n\}$. If there is no consistency between the initial and final values, a different arrangement is chosen from a mixture of $\{\textbf{A}^0_n\}$ and $\{\textbf{A}_n\}$ for the next iteration. The process is performed until numerical self-consistency is reached. The magnetic structure obtained after this iterative procedure minimizes $\mathcal{G}$, but it does not necessarily correspond to its lowest minimum. Therefore, the construction of the magnetic phase diagram requires a comparison of the Gibbs free energy of the various magnetic structures of interest. The magnetic structure with the lowest value of $\mathcal{G}$ is considered as the most stable phase for each ($\beta$,$\textbf{H}$) point. In particular, in the manuscript we have considered ferromagnetic (FM), and helical antiferromagnetic (HAFM), fan, helifan, and conical (CON) structures with a periodicity prescribed by 10 layers. The appropriate initial $\{\textbf{A}^0_n\}$ arrangement needs to be chosen carefully for the exploration of each magnetic phase.

It is important to make the following point. Our DLM-DFT theory permits a direct calculation of $\nabla_{\textbf{m}_n}\bar{\Omega}$ at each step of the iterative procedure. However, for each magnetic phase the calculation of $\mathcal{G}$ requires an evaluation of the internal energy $\bar{\Omega}$ also. We achieve this by finding an analytical expression for $\bar{\Omega}$, i.e., Eq.\ (1) in the manuscript, that fits satisfactorily our numerical DLM-DFT data of $\nabla_{\textbf{m}_n}\bar{\Omega}$. The quadratic and quartic $\mathcal{J}_{nn'}$ and $\mathcal{K}_{nn',n''n'''}$ coefficients are extracted as the result of this fitting. In the following section we give detailed information about the form of $\bar{\Omega}$ and $\nabla_{\textbf{m}_n}\bar{\Omega}$.

\subsubsection{Analytical expression for $\bar{\Omega}$}
\label{fitting}

The magnetic structures formed by the hcp HRE elements are composed of layers with FM alignment within each layer stacked perpendicular to the c-axis. The HAFM structure observed in Tb, Dy, and Ho is modulated along the c-axis and incommensurate with the lattice. The corresponding wave vector associated with the HAFM structure observed in our results (see Fig.\ 3 in the manuscript) is close to $\textbf{q}_\text{nest}\simeq$0.2$\frac{2\pi}{c}\hat{c}$, which corresponds to turn angles of about 36$^{\circ}$ in good agreement with experiment~\cite{Dobrich1}. 
To study the period of the HAFM phase we have divided the hcp lattice system into 10 layer stacks. The magnetic order of each of these ferromagnetic layers is specified by the order parameters $\{\textbf{m}_n\}$. Consequently, our internal energy $\bar{\Omega}$ only depends on the 10 different vectors $\{\textbf{m}_n, n=1,..,10\}$ that are repeated along the c-axis. Taking into account the symmetry of the lattice system, we have found the following expression to fit satisfactorily the DLM-DFT data:
%%%%%%%%%%%%%%%%%%%%%%%%%%%%%%%%%%%%%%%%%%%%%%%%%%%%%%%%%%%%%%%%%%%%%%%%%%%%%%%%%%%
\begin{equation}
\label{FittingO}
\begin{split}
\bar{\Omega}= & -\sum_{n=1}^{10}\Bigl\{\left(J_1m_n^2+K_1m_n^4\right)-\sum_{n'=1}^{5}\Bigl[\frac{1}{2}J_{1+n'}\textbf{m}_{n}\cdot(\textbf{m}_{n+n'}+\textbf{m}_{n-n'}) \\
           % + & K'_{n'}m_n^2(m_{n+n'}^2+m_{n-n'}^2)+K''_{n'}\left(\textbf{m}_{n}\cdot(\textbf{m}_{n+n'}+\textbf{m}_{n-n'})\right)^2 \\
             & +K_{n'+1}(\textbf{m}_{n}\cdot\textbf{m}_{n+n'}+\textbf{m}_{n}\cdot\textbf{m}_{n-n'})(m_n^2+m_{n+n'}^2+m_{n-n'}^2)\Bigr] \\
             & -\frac{1}{4}K'\Bigl[(\textbf{m}_{n}\cdot\textbf{m}_{n+1})(\textbf{m}_{n+2}\cdot\textbf{m}_{n+3}+\textbf{m}_{n-1}\cdot\textbf{m}_{n-2}) \\
                       & \,\,\,\,\,\,\,\,\,\,\,\,\,\,\, +(\textbf{m}_{n}\cdot\textbf{m}_{n-1})(\textbf{m}_{n-2}\cdot\textbf{m}_{n-3}+\textbf{m}_{n+1}\cdot\textbf{m}_{n+2})\Bigr]\Bigl\}.
\end{split}
\end{equation}
%%%%%%%%%%%%%%%%%%%%%%%%%%%%%%%%%%%%%%%%%%%%%%%%%%%%%%%%%%%%%%%%%%%%%%%%%%%%%%%%%%%

Hence, we have structured the coefficients $\mathcal{J}_{nn'}$ and $\mathcal{K}_{nn',n''n'''}$ in the manuscript by the 6 constants $\{J_{n},\,n=1,..,6\}$ and the 7 constants $\{K_{n}, K'',\, n=1,..,6\}$ respectively. To corroborate the accuracy of the $\mathcal{J}_{nn'}$ constants we have additionally applied a linear response theory at the PM state and recalculated them in the reciprocal space~\cite{Gyorffy1,Hughes1,IanHughes}. We verified that the PM state is unstable to the formation of a HAFM structure prescribed by a 10 layer periodicity. The long-range constants ($n-n'>6$) shown in the inset of Fig.\ 3 in the manuscript have been calculated within this approach.

We have obtained DLM-DFT data for the Gd prototype within this framework. As explained in the manuscript, we have carried out calculations for Gd with the appropriate lattice spacings to mimic Tb, Dy, and Ho~\cite{Hughes1}. The FM, HAFM, fan, helifan, and CON arrangements mentioned above provided the $\{\textbf{m}_n\}$ values for the $\nabla_{\textbf{m}_n}\bar{\Omega}$ calculations and ultimately the fitting coefficients for a given hcp $a$ and $c$ pair. We have explored the values of $A_n$=0, 0.1, 0.5, 1.0, 1.5, 2.0, 3.0, 5.0, 7.0, and 10 for each of the five magnetic structures. This corresponds to the study of order parameters with magnitudes ranging from $m_n$= 0 to 0.9. Importantly, we also sampled additional arrangements where only two layers have non-zero $A_n$ values separated by several distances and where the angles of the magnetisation directions of the two layers are also varied. We have found this to be a crucial step to be able to determine the correct form of the expression for $\Omega$.
The total number of magnetic configurations, each of them prescribed by the 10 vectors $\{\textbf{m}_n,\,n=1,..,10\}$, used for each lattice spacing was 95. The DLM-DFT data used to fit these 17 constants was totally composed, therefore, of the corresponding Weiss fields $\{\textbf{h}'_n=\nabla_{\textbf{m}_n}\bar{\Omega},\,n=1,..,10\}$ calculated at each site for each configuration.

%After some algebra it can be shown that $-\nabla_{\textbf{m}_n}\bar{\Omega}$ can be expressed as
%%%%%%%%%%%%%%%%%%%%%%%%%%%%%%%%%%%%%%%%%%%%%%%%%%%%%%%%%%%%%%%%%%%%%%%%%%%%%%%%%%%
%\begin{equation}
%\label{Fittingh}
%\begin{split}
% & -\nabla_{\textbf{m}_n}\bar{\Omega}= 2J_1\textbf{m}_n
%                +4K_1\textbf{m}_{n}m_n^2
%                +\sum_{n'=1}^{10}J_{1+n'}\left(\textbf{m}_{n+n'}+\textbf{m}_{n-n'}\right) \\
%             %   + & 2K'_{n'}\textbf{m}_{n}(m_{n+n'}^2+m_{n-n'}^2)
%             %   +    K''_{n'}\left[\textbf{m}_{n+n'}(\textbf{m}_{n}\cdot\textbf{m}_{n+n'})+\textbf{m}_{n-n'}(\textbf{m}_{n}\cdot\textbf{m}_{n-n'})\right] \\
%                + & \sum_{n'=1}^{5} K'_{n'}\Bigl(2\textbf{m}_{n}\Bigl[2\textbf{m}_{n}\cdot(\textbf{m}_{n+n'}+\textbf{m}_{n-n'})+\textbf{m}_{n+n'}\cdot\textbf{m}_{n+n'+1}+\textbf{m}_{n-n'}\cdot\textbf{m}_{n-n'-1}\Bigr] \\
%                + & \textbf{m}_{n+n'}(2m_n^2+2m_{n+n'}^2+m_{n+n'+1}^2+m_{n-n'}^2)+\textbf{m}_{n-n'}(2m_n^2+2m_{n-n'}^2+m_{n-n'-1}^2+m_{n+n'}^2)\Bigr) \\
%                + & K''\Bigl[\textbf{m}_{n+1}\left(\textbf{m}_{n+2}\cdot\textbf{m}_{n+3}+\textbf{m}_{n-1}\cdot\textbf{m}_{n-2}\right)
%                   +\textbf{m}_{n-1}\left(\textbf{m}_{n-2}\cdot\textbf{m}_{n-3}+\textbf{m}_{n+1}\cdot\textbf{m}_{n+2}\right)\Bigr].
%\end{split}
%\end{equation}
%%%%%%%%%%%%%%%%%%%%%%%%%%%%%%%%%%%%%%%%%%%%%%%%%%%%%%%%%%%%%%%%%%%%%%%%%%%%%%%%%%%

In terms of lattice Fourier transforms Eq.\ (1) of the manuscript can be written as
%%%%%%%%%%%%%%%%%%%%%%%%%%%%%%%%%%%%%%%%%%%%%%%%%%%%%%%%%%%%%%%%%%%%%%%%%%%%%%%%%%%
\begin{equation}
\label{OmFT}
\begin{split}
\bar{\Omega}= & -N\sum_\textbf{q}\left(\mathcal{J}(\textbf{q})+\sum_{\textbf{q}'}\mathcal{K}(\textbf{q},\textbf{q}')\left(\textbf{m}(\textbf{q}')\cdot\textbf{m}(-\textbf{q}')\right)\right)\left(\textbf{m}(\textbf{q})\cdot\textbf{m}(-\textbf{q})\right) \\
            = & -\sum_{\textbf{q},nn'}\left(\mathcal{J}_{nn'}+\sum_{\textbf{q}',n''n'''}\mathcal{K}_{nn',n''n'''}e^{-i\textbf{q}'\cdot(\textbf{R}_{n''}-\textbf{R}_{n'''})}\left(\textbf{m}(\textbf{q}')\cdot\textbf{m}(-\textbf{q}')\right)\right)e^{-i\textbf{q}\cdot(\textbf{R}_{n}-\textbf{R}_{n'})}\left(\textbf{m}(\textbf{q})\cdot\textbf{m}(-\textbf{q})\right),
\end{split}
\end{equation}
%%%%%%%%%%%%%%%%%%%%%%%%%%%%%%%%%%%%%%%%%%%%%%%%%%%%%%%%%%%%%%%%%%%%%%%%%%%%%%%%%%%
which directly defines some effective pair interactions in the presence of long-range magnetic order $\textbf{m}(\textbf{q}')$, i.e., shows mode-mode coupling,
%%%%%%%%%%%%%%%%%%%%%%%%%%%%%%%%%%%%%%%%%%%%%%%%%%%%%%%%%%%%%%%%%%%%%%%%%%%%%%%%%%%
\begin{equation}
\label{Effect}
\mathcal{J}^{eff.}_{\textbf{q}',nn'}=\mathcal{J}_{nn'}+\sum_{n''n'''}\left(\mathcal{K}_{nn',n''n'''}e^{-i\textbf{q}'\cdot(\textbf{R}_{n''}-\textbf{R}_{n'''})}\left(\textbf{m}(\textbf{q}')\cdot\textbf{m}(-\textbf{q}')\right)\right).
\end{equation}
%%%%%%%%%%%%%%%%%%%%%%%%%%%%%%%%%%%%%%%%%%%%%%%%%%%%%%%%%%%%%%%%%%%%%%%%%%%%%%%%%%%

When $\textbf{q}'=\textbf{0}$, i.e.\ $\textbf{m}(\textbf{0})=\textbf{m}_\text{FM}$, $\mathcal{J}^{eff.}_{nn'}=\mathcal{J}_{nn'}+\sum_{n''n'''}\mathcal{K}_{nn',n''n'''}m^2_\text{FM}$ as mentioned in the manuscript. In the language of Eq.\ (\ref{FittingO}) these effective quadratic coefficients are expressed as
%%%%%%%%%%%%%%%%%%%%%%%%%%%%%%%%%%%%%%%%%%%%%%%%%%%%%%%%%%%%%%%%%%%%%%%%%%%%%%%%%%%
\begin{equation}
\label{effective}
\begin{split}
& \tilde{J}_1=J_1+(K_1+\sum_{n'=1}^{5}4K'''_{n'})m_{\text{FM}}^2 \\
& \tilde{J}_{n+1}=J_{n+1}+\left(6K'_{n}+2K''\right)m_{\text{FM}}^2\,\,\,\,\,\,\,\,\,\,\,\,\,\,\,\,\,\,\,\,\,\,\,\,\text{for}\,\,\,\,\{n=1,...,5\}
\end{split}
\end{equation}
%%%%%%%%%%%%%%%%%%%%%%%%%%%%%%%%%%%%%%%%%%%%%%%%%%%%%%%%%%%%%%%%%%%%%%%%%%%%%%%%%%%

The modified lattice Fourier transforms shown in Fig.\ 3 in the manuscript have been calculated from the effective pair interactions shown in the expression above.

\subsubsection{Magnetic phase diagrams of Tb, Dy, and Ho}

\begin{figure*}
\centering
(a)\includegraphics[clip,scale=0.45]{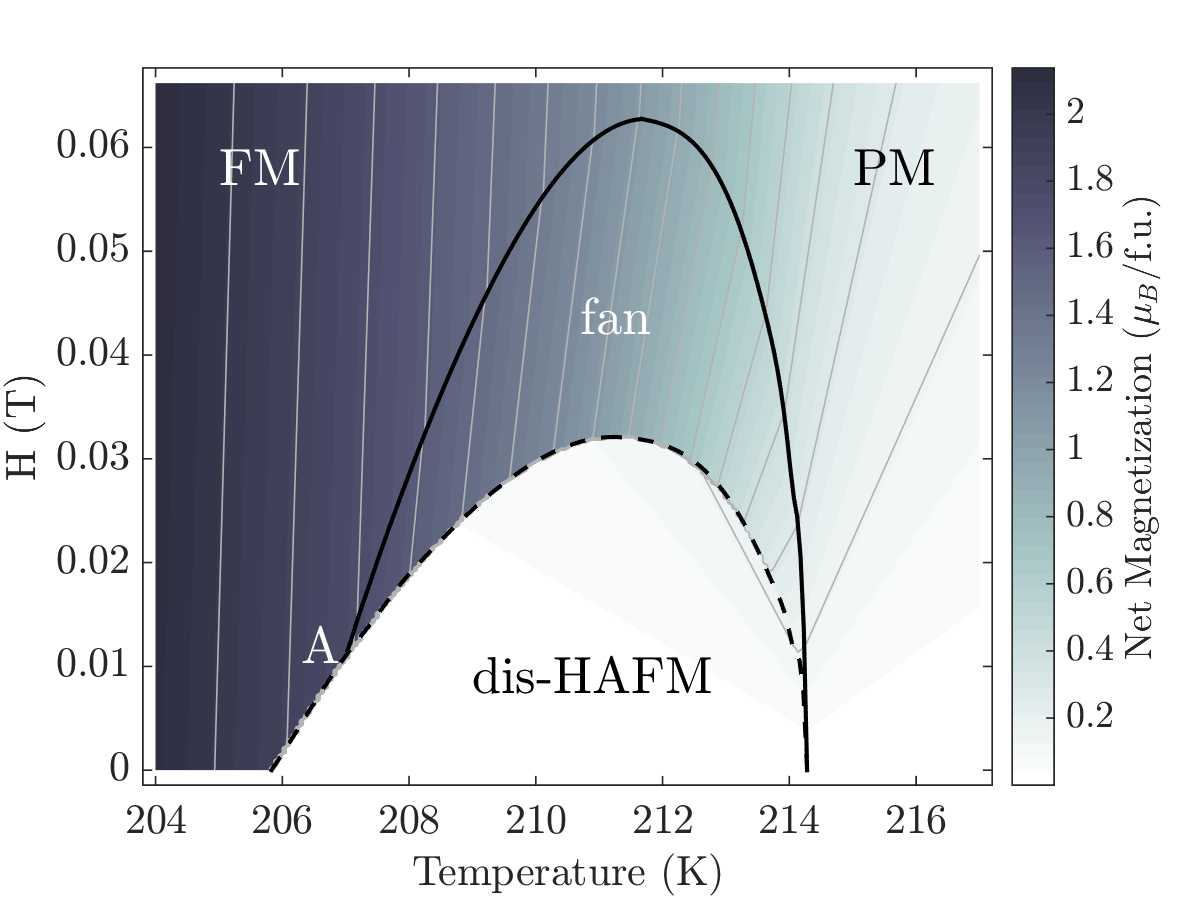}
(b)\includegraphics[clip,scale=0.45]{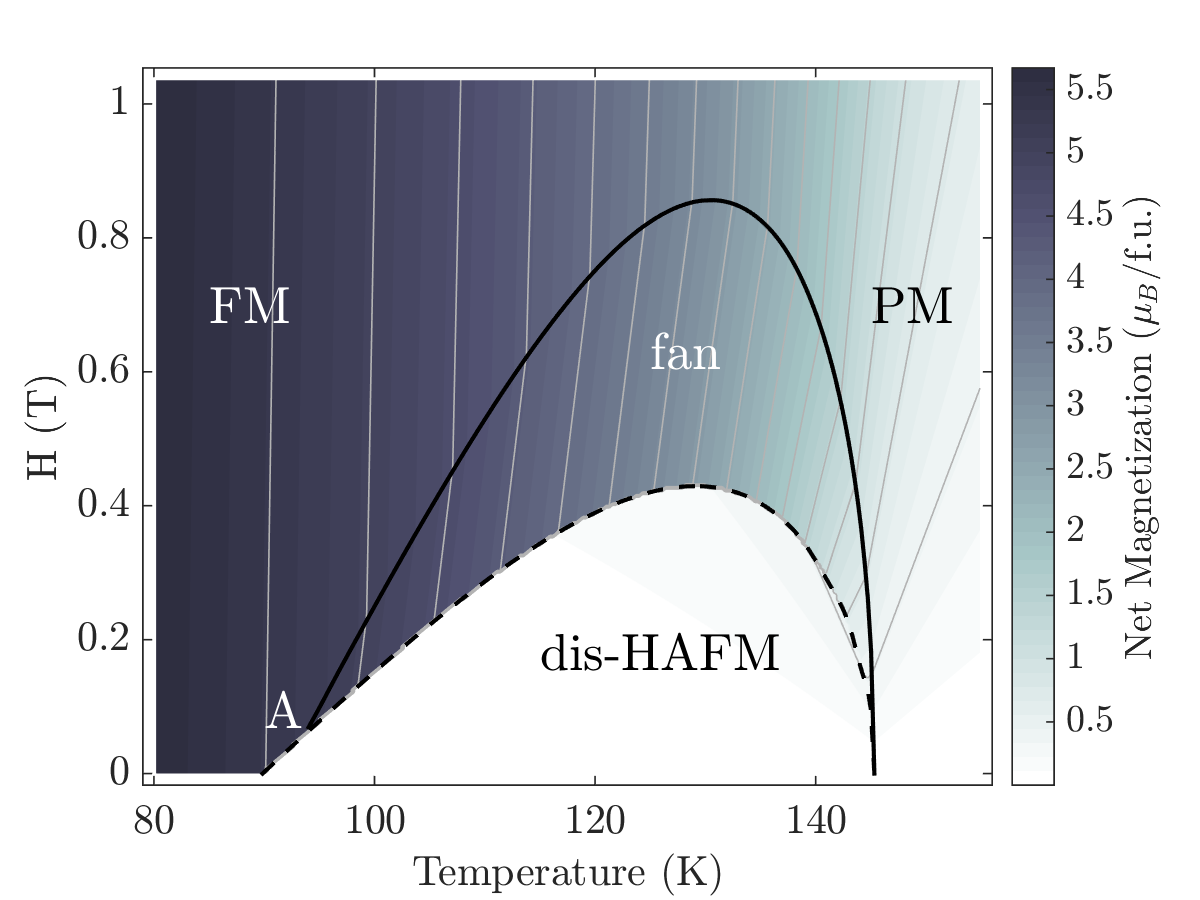}
(c)\includegraphics[clip,scale=0.45]{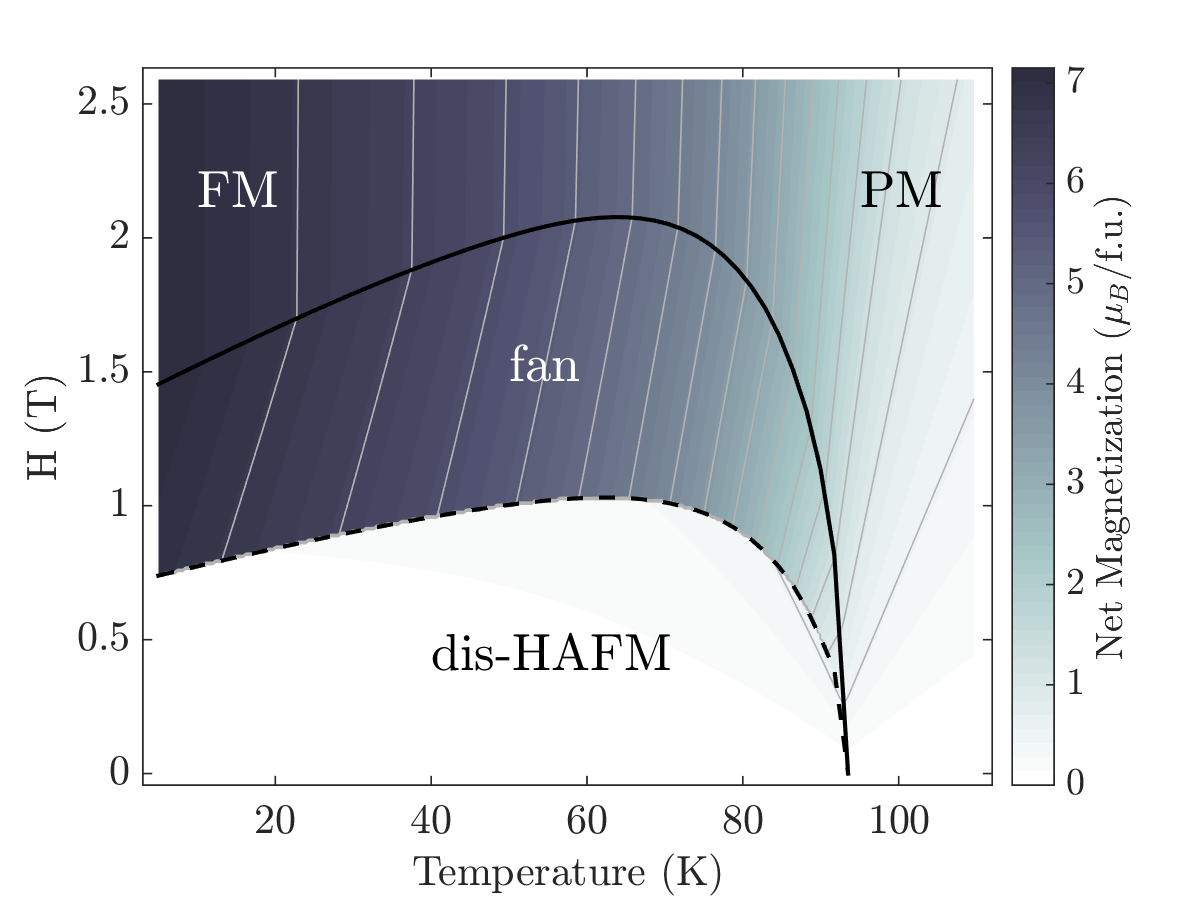}
\caption{(Color online) The magnetic phase diagram constructed for Gd with the lattice spacing of (a) Tb, (b) Dy, and (c) Ho. The de Gennes factor has been used to scale the quadratic and quartic coefficients and the magnetic field is applied along the easy direction. Continuous (discontinuous) lines correspond to second (first) order phase transitions and a tricritical point is marked (A).}
\label{PD}
\end{figure*}

We show in Table I the results of the fitting of the DLM-DFT calculation of Gd with the lattice attributes of Tb, Dy, and Ho. Fig.\ref{PD} shows the corresponding magnetic phase diagrams constructed from our results. The de Gennes factor~\cite{Mackintosh1} has been used to take into account the specific f-electron configurations. We remark that for the quartic coefficients we have used the square of the de Gennes factor. As stated in the manuscript the common features shown in Fig.\ref{PD} are determined by the mutual feedback between the magnetic ordering and the valence electrons. An inspection of Fig.\ref{PD} confirms that the lanthanide contraction favors the stability of the HAFM structure. In fact, for the lattice spacing of Ho, where the contraction is greater, the FM-HAFM transition disappears.

%%%%%%%%%%%%%%%%%%%%%%%%%%%%%%%%%%%%%%%%%%%%%%%%%%%%%%%%%%%%%%%%%%%%
\begin{table*}
\begin{center}
\begin{tabular}{ c || c c c c }
 \hline\hline
 Element & Gadolinium & Terbium & Dysprosium & Holmium \\
 \hline
% a         & 6.613  & 6.605  & 6.587  & 6.565   \\
% c         & 10.78  & 10.44  & 10.35  & 10.31   \\
% c/a       & 1.630  & 1.580  & 1.572  & 1.570   \\
 $J_1$     & 20.73  & 17.28  & 16.70  & 15.96   \\
 $J_2$     & 21.94  & 22.63  & 22.29  & 21.68   \\
 $J_3$     & 5.576  & 4.572  & 4.172  & 4.222   \\
 $J_4$     & -2.244 & -3.275 & -3.589 & -3.728  \\
 $J_5$     & 0.466  & -0.348 & -0.559 & -0.507  \\
 $J_6$     & -1.729 & -2.736 & -2.846 & -3.083  \\
 $K_1$     & 0.302  & 0.534  & 0.544  & 0.545   \\
 $K_2$    & -0.162 & -0.427 & -0.411 & -0.385  \\
 $K_3$    & -0.003 & 0.088  & 0.123  & 0.124   \\
 $K_4$    & 0.303  & 0.337  & 0.333  & 0.339   \\
 $K_5$    & -0.187 & -0.095 & -0.094 & -0.092  \\
 $K_6$    & 0.154  & 0.177  & 0.179  & 0.184   \\
 $K'$     & 0.177  & 1.020  & 1.118  & 1.231   \\

 \hline\hline
\end{tabular}
\caption{The table shows the pair ($\{J_n\}$) and quartic ($\{K_n\}$, $K'$) coefficients (in meV/f.u.) obtained from the fitting of the DLM-DFT data of Gd as a magnetic prototype of Tb, Dy, and Ho.}
\end{center}
\label{Tab11}
\end{table*}

%%%%%%%%%%%%%%%%%%%%%%%%%%%%%%%%%%%%%%%%%%%%%%%%%%%%%%%%%%%%%%%%%%%%
%%%%%%%%%%%%%%%%%%%%%%%FIGUREPD%%%%%%%%%%%%%%%%%%%%%%%%%%%%%%%%%%%%%

\begin{figure}
\centering
\includegraphics[clip,scale=0.57]{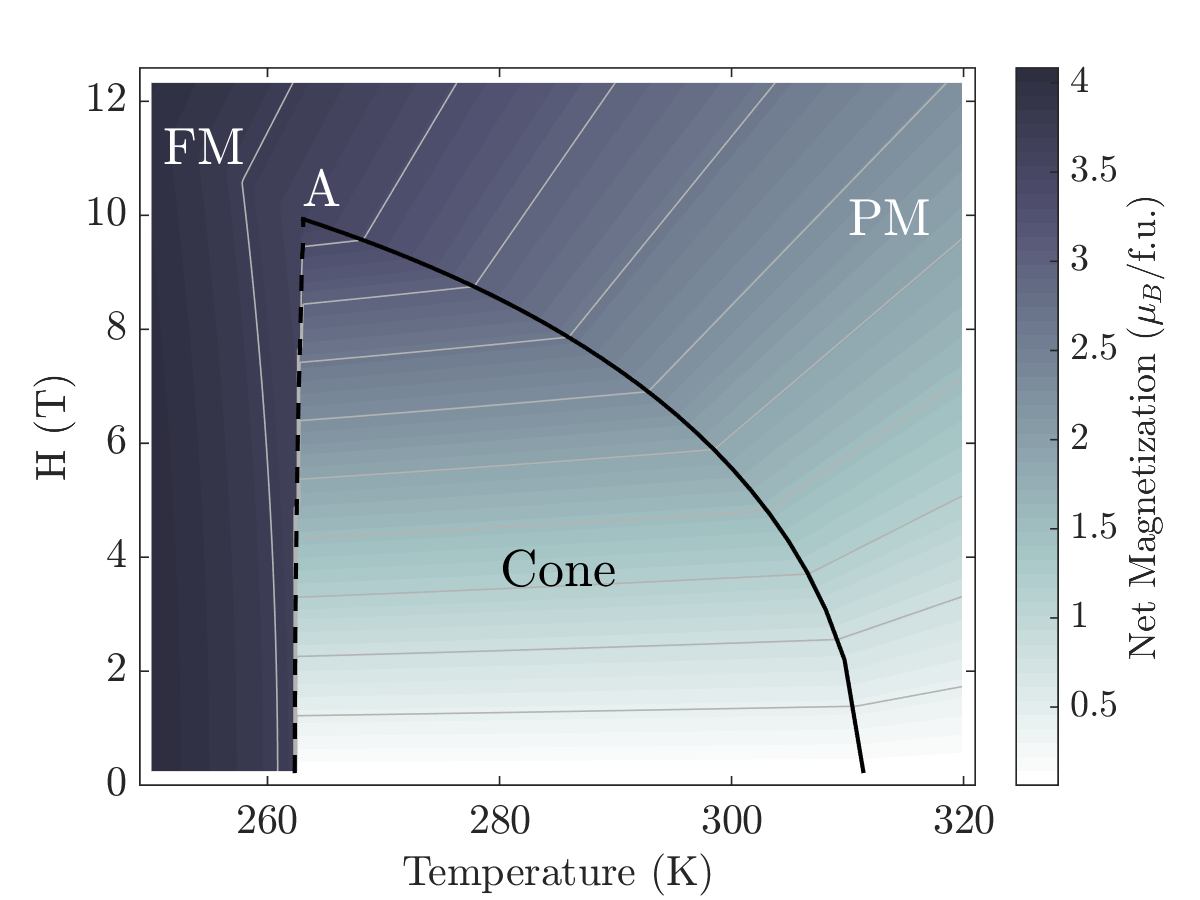}
\caption{(Color online) The magnetic phase diagram constructed for Gd with the lattice spacing of Dy when the magnetic field is applied along the hard direction. Continuous (discontinuous) lines correspond to second (first) order phase transitions and a tricritical point is marked (A).}
\label{Dy1}
\end{figure}

In Fig.\ \ref{Dy1} we show the magnetic phase diagram to accompany Fig.\ 1 in the manuscript (Dy lattice spacing) when the magnetic field is applied along the hard direction. The figure shows that the cone structure is stabilized if the magnetic field is applied parallel to the hard axis. The transition from the cone to the FM/PM state is of second (first) order at high (low) temperatures.

\begin{figure}
\centering
\includegraphics[clip,scale=0.57]{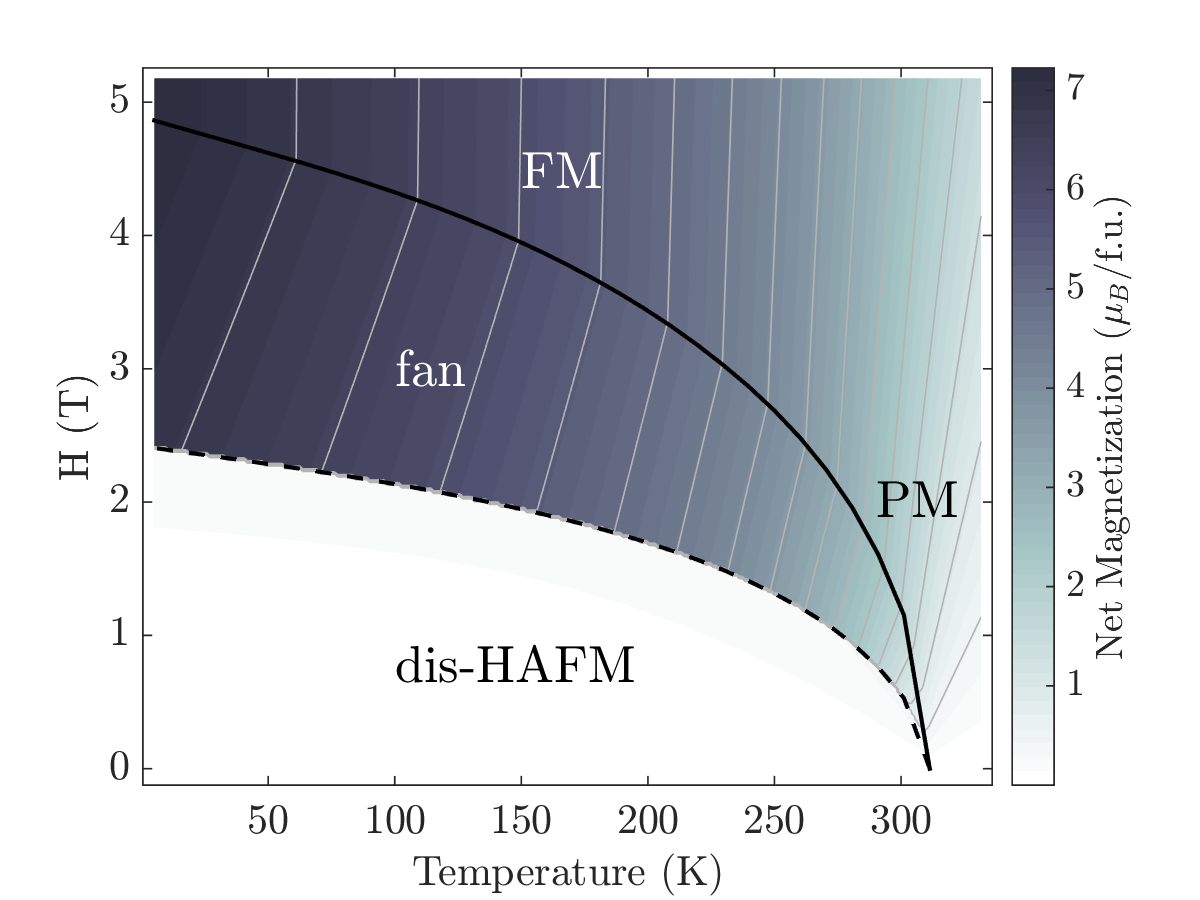}
\caption{(Color online) The magnetic phase diagram constructed for Gd with the lattice spacing of Dy when the quartic coefficients are set to zero and the magnetic field is applied along the easy direction. Continuous (discontinuous) lines correspond to second (first) order phase transitions.}
\label{Dy2}
\end{figure}

Fig. 1 in the manuscript is radically altered and qualitatively at odds with experimental results \cite{Chernyshov1} if we neglect the quartic terms, i.e. set $\mathcal{K}_{nn',n''n'''}$=0, and so omit the feedback between the valence electronic structure and lanthanide local f-electron magnetic moment order. This is shown in Fig.\ \ref{Dy2}.

%%%%%%%%%%%%%%%%%%%%%%%%%%%%%%%%%%%%%%%%%%%%%%%%%%%%%%%%%%%%%%%%%%%%
%%%%%%%%%%%%%%%%%%%%%%%FIGUREDy1%%%%%%%%%%%%%%%%%%%%%%%%%%%%%%%%%%%%
%%%%%%%%%%%%%%%%%%%%%%%%%%%%%%%%%%%%%%%%%%%%%%%%%%%%%%%%%%%%%%%%%%%%
%%%%%%%%%%%%%%%%%%%%%%%FIGUREDy2%%%%%%%%%%%%%%%%%%%%%%%%%%%%%%%%%%%%

\bibliography{/home/theory/phrsns/Documents/PhD2/PRL/PRL_2016_JS_V3/bibliography.bib}

\end{document}